\documentclass[12pt,thmsa]{article}

\usepackage{amsfonts}
\usepackage{graphicx}
\usepackage{epsfig}
\usepackage{graphics}

\makeatletter
\newcommand{\row}[1]%
{\mathord{\buildrel{\lower3pt%
\hbox{$\scriptscriptstyle\rightarrow$}}\over #1}}

\newcommand{\dyadic}[1]{\mathord{\dyadic@rrow{#1}}}
\newcommand{\dyadic@rrow}[1]{
\begin{picture}(12,12)(-1,0)
\put(-1,9){\makebox(0,0)[t]{$\scriptscriptstyle\downarrow$}}
\put(-1,9){\makebox(0,0)[l]{$\scriptscriptstyle\longrightarrow$}}
\put(5,0){\makebox(0,0)[b]{$#1$}}
\end{picture}
}

\newcommand{\bra}[1]{\bigl\langle #1 \bigr|}
\newcommand{\ket}[1]{\bigl| #1 \bigr\rangle}

\newcommand{\tottr}{^{\mathsf{T}}}

\begin{document}

\begin{center}{\Large Entanglement Routers via Wireless Quantum  Network  Based on
Arbitrary Two Qubit Systems\\} {N. Metwally\\}

$^1$Department of Mathematics,
Faculty of Science,
Aswan University, Aswan, Egypt \\
$^2$Mathematics Department, College of Science, Bahrain
University, Bahrain \\
\end{center}
\date{\today }

\begin{abstract}

A wireless quantum network is generated between multi-hop, where
each hop consists of two entangled nodes. These nodes share a
finite number of  entangled two qubit systems randomly. Different
types of   wireless quantum  bridges are generated between the
non-connected nodes.
 The efficiency of these  wireless quantum  bridges to be used as quantum channels
between its terminals to  perform quantum teleportation  is
investigated. We suggest a theoretical wireless quantum
communication protocol to teleport unknown quantum signals from
one node to  another, where the more powerful wireless quantum
bridges are  used as quantum channels. It is shown that, by
increasing the efficiency of the sources which emit the initial
partial entangled states, one can increase the efficiency of the
wireless quantum communication protocol.

\end{abstract}

PAC: 03.65.Aa, 03.65.Ud, 03.65.Yz, 03.67.Hk, 03.67.Lx

\topmargin=0.001cm \textheight=24cm \textwidth=17cm
\section{Introduction}

Communication and exchange information are the most repaid
developed phenomena. The current technologies  which are used to
transmit, store  and manipulate information are  developed each
short period of time. The most challenge of these classical
devices is the possibility of communicating and exchange
information securely \cite{Yang}. However, quantum techniques of
manipulating information are developed parallel to the classical
ones and they are more secure than the classical technology
\cite{Hemmer}. Quantum networks represent one of the most recent
developments  in the context of  quantum communications
\cite{Giraud,Oliver, Skye, Arda}. There are several types of these
networks that have introduced. For example, the possibility of
building quantum router based on ac control of qubit chains is
discussed by Zueco et al. \cite{Zueco}. Duan and Monroe
\cite{Duan} have generated quantum network with trapped ions.
Generating wireless quantum network between Josephen qubit is
investigated by Sergeenkov and Rotoli \cite{Rot}.
 Chudzicki and Strauch \cite{Stra} studied the routing of quantum
 information in parallel on multidimensional networks of tunable
 qubits and oscillators.
Spin networks have been  used  by Ross and Kay \cite{Kay} to route
quantum information perfectly. Generating quantum network between
six maximum entangled  qubits  by Dzyaloshinskii- Moriya (DM)
interaction is investigated by Metwally{\cite{Metwally2011}.
Moreover,  Abdel-Aty et al., \cite{Haleem2014} used  DM
interaction to generated quantum network between partial entangled
qubits.
 Cheng et al. \cite{Cheng} have
introduced a quantum routing  mechanism to teleport  unknown
quantum state from one quantum device to another by using their
model of the wireless wide-area network. Routing quantum
information  via $XX$ spin chain has been investigated by
Paganelli et al.\cite{Paga}. The concept of distributed wireless
quantum communication networks is considered by  Tao et al.
\cite{Yu}. Recently, Wang et al. \cite{Wang} have proposed a
scheme for faithful quantum communication in quantum wireless
multi- hop network, where they assumed that, the intermediate
nodes share arbitrary pairs of Bell states.

This motivates us to investigate the possibility of generating
wireless quantum network (WQN) between different disconnected
 hops' members. This protocol is different from the others, where we assume
that, the sending station contains  three sources the first source
$\mathcal{S}_1$,  has the ability to emit different types of
quantum signals (two-qubit systems). These quantum signals may be
maximum entangled states as Bell states \cite{Nielsen}or partial
entangled states as Werner  \cite{Werner} and $X$ \cite{Eberly}
states or generic pure states \cite{Englert}. However, to be sure
that each
 hop has at least one Werner state, the second source
$\mathcal{S}_2$ supplies all the  hops' nodes with Werner states.
The function of the third source $\mathcal{S}_3$ is supplying the
nodes with the required unknown  quantum signals to be teleported
between the different nodes.

The structure of the paper is described as follows. In  Sec. 2, we
report the suggested theoretical wireless communication  protocol.
Sec.(2.1) is devoted to the distribution of the quantum signals to
between hops'members.
 In Sec. (2.2), we describe how one can
generate different wireless quantum  bridges (QWBS) to be used as
quantum channels to perform quantum teleportation. The efficiency
of the generated WQBS to achieve quantum teleportation is
discussed in Sec.(2.3). Teleporting unknown quantum signals from
one member to another is studied in Sec.(2.4).  The concept of
purification is described shortly in Sec.3.  Finally, in Sec.(4),
we discussed our results.

\section{The suggested Protocol} As we have mentioned above,  we have three
sources. These sources are  similar to a source with multiple
antennas that transmit different entangled or separable quantum
signals  to network's members, who located in different hops. This
type of transmitter in classical context is called multiple-input
and multiple-output (MIMO), which transmit different signals.
Similarly, we called this source is QMIMO.  The following steps
summarize the suggest protocol:
\begin{enumerate}

\item {Quantum signals distribution\\}At the  sending station, one
antenna of the quantum MIMO (QIMIO) supplies the nodes in each hop
with different types of quantum signals, MES, PES, or separable
states (SS), meanwhile the second antenna supplies the other  hops
with different types of Werner states randomly. The third antenna
sends the unknown quantum signals which are needed to be
teleported from one
 hops'partner (node) to another. The details of distributing the
different quantum signals on the  hops's partners are given in
Fig.(1a).  Fig.(1b) shows the structure of the WQN clearly, where
two  hops with two nodes are considered. Each hop's nodes share a
class of partial entangled quantum signal of Werner type. Moreover
the nodes share a finite number of partial entangled quantum
signals with the other nodes \cite{Liang}.

 \item{Wireless  Quantum Bridges\\} If one  of the  hops'
partners receives a unknown quantum  signal (qubit) and he/she is
asked to send it to another member in the WQN, he/she has to
generate a wireless quantum bridge (WQB) to be used as quantum
channel.  The two nodes are called quantum neighbors, if they
share at least one of the Werner state.  However, if the two nodes
are not quantum neighbors, then the sender generates a wireless
quantum bridge with the most nearer one  to the required member.
In Fig.(2), we show how the non-connected nodes generate a
wireless quantum bridge.

\item{Bridges' efficiency\\} The wireless quantum  bridge's
partners (nodes)  check if their   WQB  has the ability to be used
as quantum channel to perform quantum teleportation or not. If
yes, they move to the second step. If not, they send it to the
purification lab to improve their  efficiency.

\item{Teleportation step\\} As soon as the WQB is generated, the
sender performs the CNOT operation and Hadamard gate between
his/her qubits followed by Bell measurements. The sender sends
his/her results to the receiver who retrieves the original state
by performing a suitable local operations. The details are given
in Sec. (2.4).

\item{Purification step\\}If the generated wireless bridges are
not efficient  to be used as quantum channels to perform
teleportation, then they are sent to the  purification lab to
increase their entanglement and hence their ability to achieve
quantum teleportation.

\end{enumerate}
In the following subsections, the previous steps of the suggested
theoretical wireless communications are investigated extensively
and we show our idea by different cases.
\begin{figure}[t!]
\begin{center}
\includegraphics[width=22pc,height=22pc]{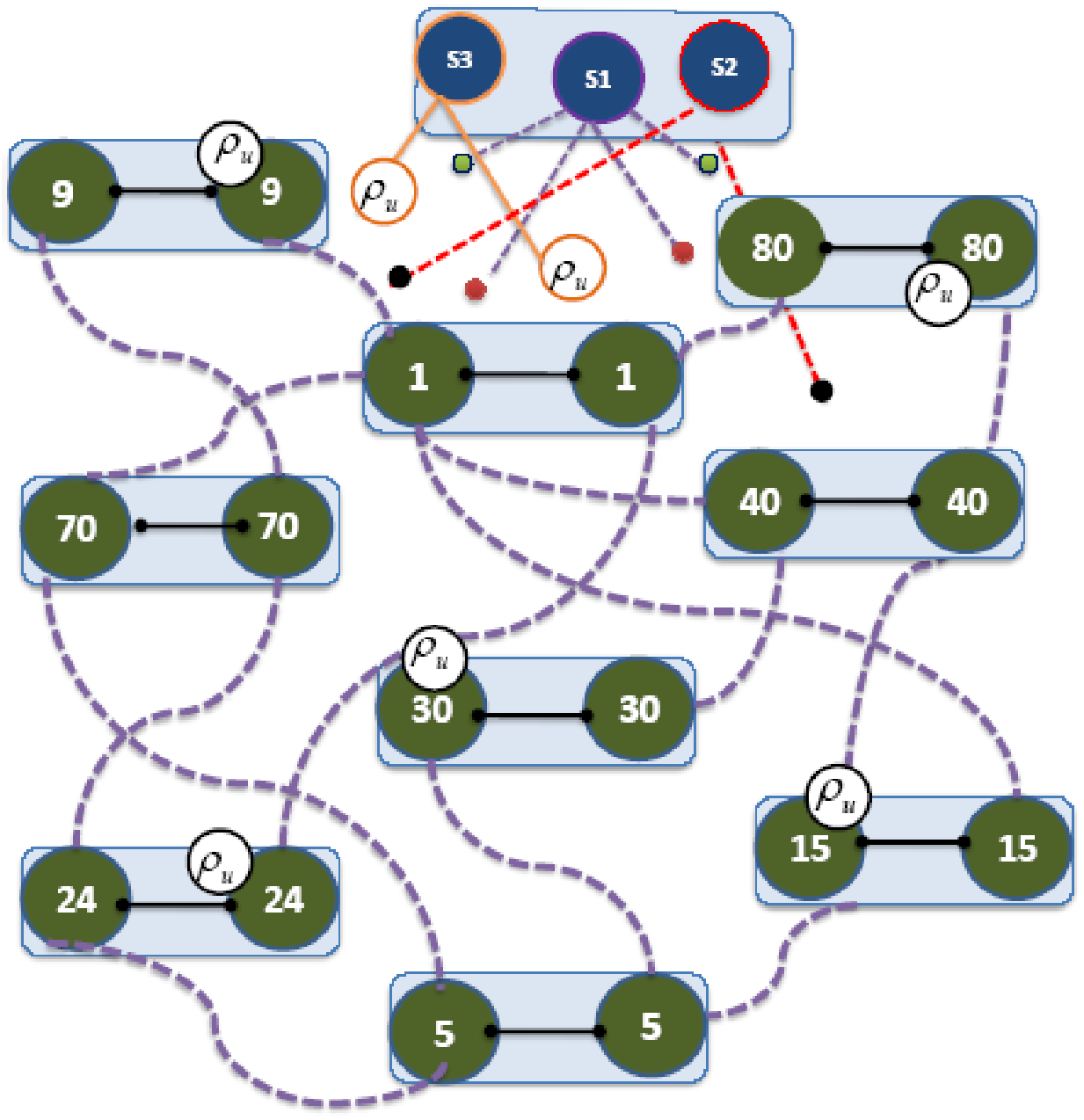}
\includegraphics[width=16pc,height=12pc]{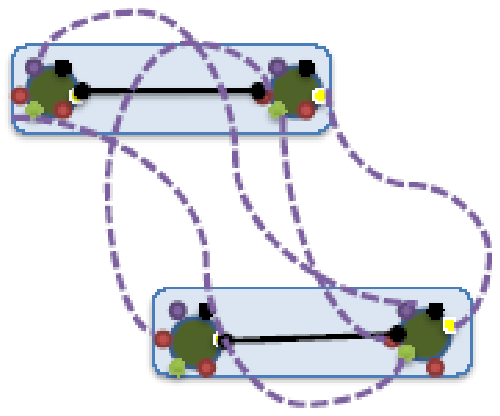}
 \put(-320,-10){$(a)$}
 \put(-90,-10){$(b)$}
\end{center}
 \caption{(a) This figure shows the distribution of the quantum signals to the
 different  hops'nodes where the numbers on the circle represent the
 hops' location. Each  hop consists of  entangled  two  nodes.
  The  source $\mathcal{S}_1$, supplies the nodes randomly with  different types of entangled quantum signals
  , Bell, $X$, pure  or Werner states (dash-curves). The source $\mathcal{S}_2$ supplies
  each   hop's node with Werner state (solid curves). The source $\mathcal{S}_3$
  supplies some nodes with unknown  quantum signals $\rho_u$ to be teleported  from one node to another
  using this wireless quantum network.
  (b) A two  hops, where each  hops's node share a class of Werner state (solid-curves).
  Each node share a finite numbers of  partial  entangled quantum signals with  other nodes. }
\end{figure}

\subsection{Quantum Signal  Distribution}
As it has mentioned above, one antenna of the QMIMO  sends
different quantum signals to the  hops'nodes. The emitted initial
quantum signals  to the  hops' nodes are classified as maximum
entangled, partial entangled or separable states. The class of
maximum entangled states (Bell states) includes
$\rho_{\phi^\pm}=\ket{\phi^\pm}\bra{\phi^\pm}$ and
$\rho_{\psi^\pm}=\ket{\psi^\pm}\bra{\psi^\pm}$, where
$\ket{\phi^\pm}=\frac{1}{\sqrt{2}}(\ket{00}\pm\ket{11})$ and
$\ket{\psi^\pm}=\frac{1}{\sqrt{2}}(\ket{01}\pm\ket{10})$. These
entangled states can be described by using Pauli operators as
\cite{Nielsen},

\begin{eqnarray}\label{iniS}
 \rho_{Bell}&=& \left\{\begin{array}{ll}
\frac{1}{4}(1+\sigma_1^{(i)}\sigma_1^{(j)}+\sigma_2^{(i)}\sigma_2^{(j)}+\sigma_3^{(i)}\sigma_3^{(j)}),& \\
\frac{1}{4}(1+\sigma_1^{(i)}\sigma_1^{(j)}-\sigma_2^{(i)}\sigma_2^{(j)}+\sigma_3^{(i)}\sigma_3^{(j)}),&  \\
\frac{1}{4}(1-\sigma_1^{(i)}\sigma_1^{(j)}+\sigma_2^{(i)}\sigma_2^{(j)}+\sigma_3^{(i)}\sigma_3^{(j)}),& \\
\frac{1}{4}(1-\sigma_1^{(i)}\sigma_1^{(j)}-\sigma_2^{(i)}\sigma_2^{(j)}-\sigma_3^{(i)}\sigma_3^{(j)}),& \\
\end{array} \right.
\end{eqnarray}
where $\sigma^{(i(j))}_{k}, k=1,2,3$ are the Pauli operators for
the qubits $"i" $ and $"j"$, respectively and
$\sigma^{((i)j)}_1=\ket{0}\bra{1}+\ket{1}\bra{0}$,
$\sigma^{((i)j)}_2=i(-\ket{0}\bra{1}+\ket{1}\bra{0})$ and
$\sigma^{((i)j)}_3=\ket{0}\bra{0}-\ket{1}\bra{1}$. It is clear
that, any one of these states can be transferred into another one
by using local operations. The second types of the transmitted
states from the QMIMO  are partial entangled states. In this
contribution, we consider two classes: $X$ and generic pure
states, which can be defined as,

\begin{eqnarray}\label{iniS}
 \rho_{PES}&=& \left\{\begin{array}{ll}
\rho_{pure}=\frac{1}{4}\{1+p(\sigma_1^{(i)}-\sigma_1^{(j)})-\sigma_1^{(i)}\sigma_1^{(j)}-q(\sigma_2^{(i)}\sigma_2^{(j)}+\sigma_3^{(i)}\sigma_3^{(j)})\},& \\
\rho_{X}=\frac{1}{4}\{1-c^{(ij)}_{11}\sigma_1^{(i)}\sigma_1^{(j)}-c^{(ij)}_{22}\sigma_2^{(i)}\sigma_2^{(j)}-
c^{(ij)}_{33}\sigma_3^{(i)}\sigma_3^{(j)}\},&  \\
\end{array} \right.
\end{eqnarray}
where $p=\sqrt{1-q^2}$. It is clear that, if we set $p=0$ in
$\rho_{pure}$,  one gets a maximum entangled state (fourth state
(Eq.(1)).  This entangled  pure state turns into a separable state
if we set $p=1$. However, this type of the pure states can be
transformed into four equivalent forms by local operations and
consequently all the maximum entangled states can be obtained from
the other forms of these pure states \cite{Englert}. The degree of
entanglement of this pure state is given by   Wootter's
concurrence \cite{Wootters} as,
\begin{equation}
\mathcal{C}_{pure}=max\Bigl\{0,\frac{1}{2}(1+2q)-\frac{1}{2}\Bigr\}.
\end{equation}
The second type of the transmitted partial entangled states is
called $X-state$ \cite{Eberly}, where one can obtain what is
called Werner state \cite{Werner}, $\rho_{w}$ by setting
$c_{11}=c_{22}=c_{33}=x$ and $\rho_{Bell}$ if we set $x=1$.  The
degree of entanglement of the  $X-$  state
 is given by,

\begin{equation}
\mathcal{C}_X=max\Bigl\{0,
\frac{3}{2}tr\{\dyadic{C}\tottr{\dyadic{C}}\}-\frac{1}{2}\Bigr\},
\end{equation}
where $\dyadic{C}$ is a $3\times 3$ matrix called cross dyadic
represents the correlation between the two qubits. The  non-zero
elements of the cross dyadic $\dyadic{C}$ are given by $c_{11},
c_{22}$ and $c_{33}$.

\begin{figure}
\begin{center}
\includegraphics[width=25pc,height=15pc]{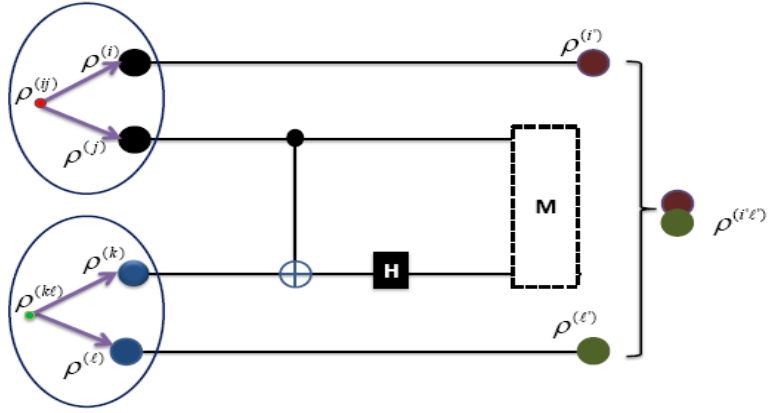}
\end{center}
 \caption{Circuit for generating wireless bridges between different  hops.}
\end{figure}

\subsection{Wireless Quantum Bridges:Entanglement Routing}
Now, each node in  different  hops has its  own qubits. The aim of
this section is generating  wireless quantum bridges (WQBS)
between any two non-connected nodes  located in different hops.
This procedure can be achieved via CNOT operation and Hadamard
gate followed by Bell measurements \cite{Cheng} as shown in
Fig.(2).

Let us first consider two  hops their partners share  a class of
$X-$ state, where the first  hop's nodes  share the state
$\rho^{(ij)}$ while the second  hop's nodes share the state
$\rho^{(k\ell)}$. The nodes $(j)$ and $(k)$ perform CNOT operation
 followed by  Hadamard gate on
the qubit $(k)$. After performing  Bell measurements on the qubits
$(j\&k)$, the final state is projected into $\rho^{(i'\ell')}$
(see Fig.(2)). The final state $\rho^{(i'\ell')}$ represents the
generated wireless quantum bridge between the nodes $(i\& j)$.
However, if the nodes of the first and the second  hops share $X-$
states, then we call the generated entangled states  by $XX$
wireless bridge. In the computational basis, $00,01,10, 11$, this
bridge can be written as,

\begin{equation}
\rho_{XX}= \left(\begin{array}{cccc}
\varrho_{11}&0&0&\varrho_{14}\\
0&\varrho_{22}&\varrho_{23}&0\\
0&\varrho_{32}&\varrho_{33}&0\\
\varrho_{41}&0&0&\varrho_{44}
\end{array}
\right),
\end{equation}
where
\begin{eqnarray}
\varrho_{11}&=&\frac{1}{2}(\mathcal{A}_1\mathcal{B}_1+\mathcal{A}_2\mathcal{B}_2),
\quad
\varrho_{14}=\frac{1}{2}(\mathcal{A}_3\mathcal{B}_3+\mathcal{A}_4\mathcal{B}_4),
\nonumber\\
\varrho_{22}&=&\frac{1}{2}(\mathcal{A}_1\mathcal{B}_2+\mathcal{A}_2\mathcal{B}_1),
\quad
\varrho_{23}=\frac{1}{2}(\mathcal{A}_3\mathcal{B}_4+\mathcal{A}_4\mathcal{B}_3),
\nonumber\\
\varrho_{32}&=&\varrho_{23},\quad \varrho_{33}=\varrho_{22},\quad
\varrho_{41}=\varrho_{14},\quad \varrho_{44}=\varrho_{11},\quad
\mbox{with} \nonumber\\
 \mathcal{A}_1&=&\mathcal{B}_1=\frac{1+c_{33}}{4},\quad
 \mathcal{A}_2=\mathcal{B}_2=\frac{1-c_{33}}{4},
\nonumber\\
 \mathcal{A}_3&=&\mathcal{B}_3=\frac{c_{11}-c_{22}}{4},\quad
  \mathcal{A}_4=\mathcal{B}_4=\frac{c_{11}+c_{22}}{4}.
\end{eqnarray}
From this wireless quantum bridge, one can  obtain the following
bridges:
\begin{enumerate}
\item If we set $c^{(ij)}_{xx}=c^{(ij)}_{yy}=c^{(ij)}_{zz}=x$ and
$c^{(k\ell)}_{11}=c^{(k\ell)}_{22}=c^{(k\ell)}_{33}=x$, one gets
the  wireless Werner-Werner  quantum bridge ($WW$-bridge).

\item If we set $c^{(ij)}_{11}=c^{(ij)}_{22}=c^{(ij)}_{33}=x$ and
$c^{(k\ell)}_{11}=c^{(k\ell)}_{22}=c^{(k\ell)}_{33}=-1$, one gets
the  wireless Werner-Bell  quantum bridge $WB$-bridge.

\item If we set $c^{(ij)}_{11}=c^{(ij)}_{22}=c^{(ij)}_{33}=x$  and
$c^{(k\ell)}_{11}\neq c^{(k\ell)}_{22}\neq c^{(k\ell)}_{33}\neq 0$
one gets the wireless Werner-$X$  quantum bridge $WX$-bridge.
\end{enumerate}
 However, if the node of  one  hop
share initially $X$ state while the nodes of the another  hop
share a class of the generic pure state, then the  generated
wireless quantum bridge is called $XP$ bridge. In the
computational basis, this bridge can be described by a density
matrix  of size $4\times 4$, its elements are given by,
\begin{eqnarray}
\tilde\varrho_{11}&=&\mathcal{A}_1\mathcal{C}_1+\mathcal{A}_2\mathcal{C}_2,
\quad
\tilde\varrho_{12}=-(\mathcal{A}_1\mathcal{C}_3+\mathcal{A}_2\mathcal{C}_3),
\quad
\tilde\varrho_{13}=\mathcal{C}_3(\mathcal{A}_3+\mathcal{A}_4),
\nonumber\\
\tilde\varrho_{14}&=&(\mathcal{A}_3\mathcal{C}_1+\mathcal{A}_4\mathcal{C}_2),
\quad \tilde\varrho_{21}=\tilde\varrho_{21},\quad
\tilde\varrho_{22}=\mathcal{A}_1\mathcal{C}_2+\mathcal{A}_2\mathcal{C}_1,
\quad
\tilde\varrho_{23}=\mathcal{C}_2(\mathcal{A}_3+\mathcal{A}_4),\quad
\nonumber\\
 \tilde\varrho_{24}&=&\tilde\varrho_{13}\quad
 \tilde\varrho_{23}=\tilde\varrho_{32}, \quad
 \tilde\varrho_{33}=\tilde\varrho_{22},\quad
 \tilde\varrho_{34}=\tilde\varrho_{12},\quad
 \tilde\varrho_{41}=\tilde\varrho_{14},\quad \tilde\varrho_{42}=\tilde\varrho_{24}
 \nonumber\\
 \tilde\varrho_{43}&=&\tilde\varrho_{34},\quad
 \tilde\varrho_{44}=\tilde\varrho_{11},
\end{eqnarray}
where $\mathcal{A}_i, i=1..4$ are given from (6) and
$\mathcal{C}_1=\frac{1-q}{4}, \quad \mathcal{C}_2=\frac{1+q}{4}$
and $\mathcal{C}_3=\frac{p}{4}$.

 Fig.(3a) describes the behavior of the degree of entanglement
between the terminals of the  generated wireless quantum bridge
(WQBS). {\it Firstly}, we assume that, the non-connected nodes
share a  $WW$ bridge.  From this figure, we can see that the
entanglement between the terminals of
 $WW$ bridge  is generated  at  $x>0.578$. However, for larger
values of $x$, the entanglement between the terminals of $WW$
bridge increases and reaches its maximum values, i.e.,
($\mathcal{E}=1$ at $x=1$, namely, the initial states are Bell
states). {\it Secondly}, the nodes of the first  hop share a class
of Werner type, while the second  hop's nodes share a class of
$X-$ state which is defined by $c_{11}=-0.9, c_{22}=-0.8$ and
$c_{33}=-0.7$. In this case, the entanglement between the
terminals of the wireless quantum bridge $(WX$) is generated at
$x>41$. As $x$ increases, the entanglement increases  to reach its
maximum value $(\mathcal{E}=0.7$ at $x=1)$. Thirdly, the nodes of
one  hop share a maximum entangled state (Bell types), while the
second  hop's nodes share Werner state. In this case, the
entanglement between the $WB$ bridge terminals is generated for
smaller values of $x(=0.34)$.

\begin{figure}[t!]
\begin{center}
\includegraphics[width=20pc,height=15pc]{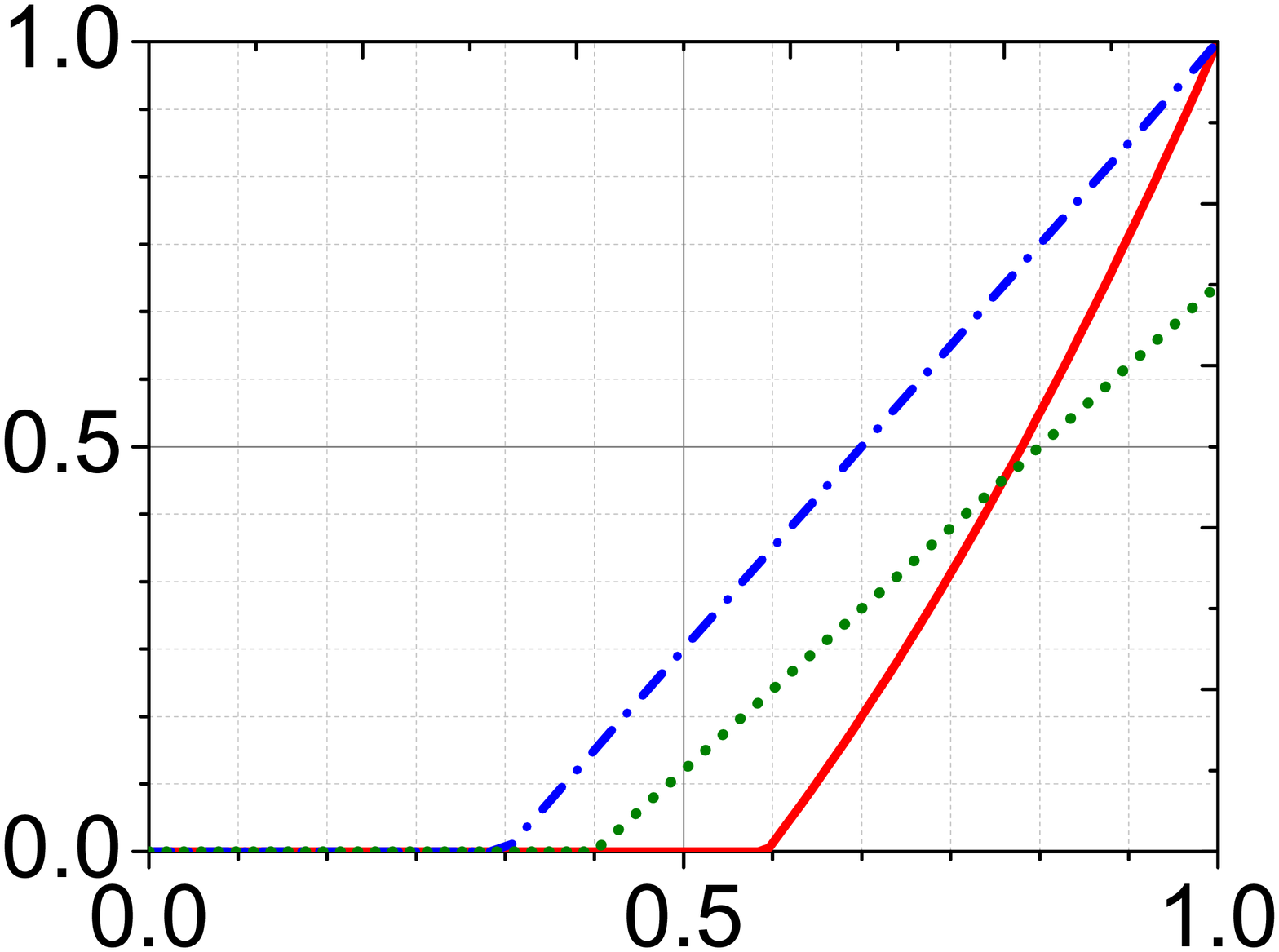}
\put(-120,0){\Large $x$}
 \put(-245,90){\Large$\mathcal{C}_{\mu\nu}$}
 \put(-70,150){$(a)$}
 \includegraphics[width=20pc,height=15pc]{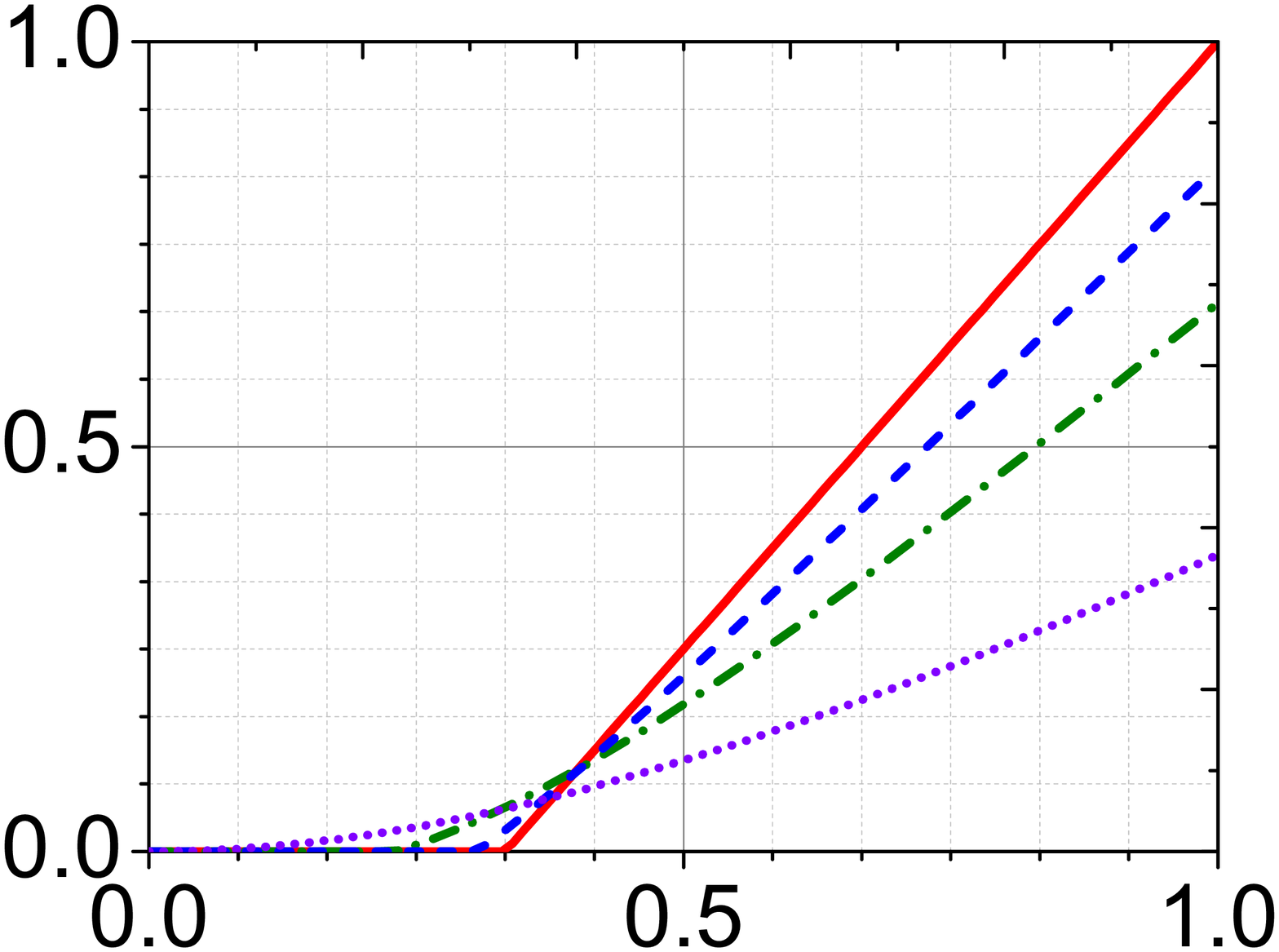}
 \put(-120,0){\Large $x$}
 \put(-250,88){\Large $\mathcal{C}_{WP}$}
  \put(-60,150){$(b)$}
\end{center}
\caption{The entanglement $\mathcal{C}_{\mu\nu}$, of the  wireless
quantum bridges between the non-interacted  hops. (a)The solid,
dot and dash-dot curves for $WW,WX$ and $WB$, respectively. (b)
The entanglement between the terminales of the wireless quantum
bridge $(WP)$, where  the solid, dash, dash-dot and dot curves for
$p=0,0,0.9,0.7$ and $1$, respectively.}
\end{figure}

In Fig.(3b), we assume that, the  hops share two different initial
quantum signals. The first  hop's nodes share Werner sate, while
the nodes of the second  hop share a class of pure state. However,
if the  second  hop is supplied with different initial entangled
pure states, where we set $p=0,0.7,0.9, 1$. It is clear that, at
$p=0$, which corresponding to Bell state, the two  hops entangle
together at $x=0.33$. The degree of entanglement between the
terminals of the wireless $WP$ bridge increases as $x$ increases
to reach its maximum value ($\mathcal{E}=1$ at $x=1)$, where the
initial two quantum signals  are maximum entangled states. As one
increases $p$, namely the second  hop is supplied with less
entangled pure state, the entanglement between the two
 hops appears suddenly  for smaller values of $x$. However, the
maximum values of the entanglement is reached at $x=1$, where it
is smaller than "1" for larger values of $p$. Starting from
separable state, where we set $p=1$, the two  hops generate  a
wireless quantum bridge  at very small value of $x$, but the
degree of entanglement between the bridge' terminals is very small
compared with those depicted for entangled pure state.

From  Fig.(3), we can conclude that, it is possible to entangle
different  hops, their partners share arbitrary classes of initial
two qubit systems. The results show that, if each  hop's nodes
share a pure states even they are initially  separable, one can
generate entangled wireless quantum bridges between the hops'
nodes. Using Werner state with larger value of its parameter
($x>0.5$), one can generate wireless quantum bridges between the
 hops' nodes with high degree of entanglement.

\subsection{Bridges efficiency  }
 In this section, we investigate the efficiency of the generated
wireless quantum bridges (WQBS), where we discuss the possibility
of  using them as quantum channels  to perform quantum
teleportation.  The inequality which measures the efficiency of
the   WQBS
 to perform quantum teleportation is given by \cite{Horodecki},
 \begin{equation}
 Telp=tr\{\sqrt{\dyadic{C}^{\tottr}\cdot\dyadic{C}}\}>1,
 \end{equation}
where the elements of the cross dyadic  $\dyadic{C}$  are given by
$c_{mn}=tr\{\sigma^{(1)}_m\sigma^{(2)}_n\rho_{B}\}, m,n=1,2,3$ and
$\rho_{B}$ stands for the state of the wireless quantum bridge.
For example, $c_{11}=tr\{\sigma_1^{(1)}\sigma_1^{(2)}\rho_{B}\}$,
$c_{12}=tr\{\sigma_1^{(1)}\sigma_2^{(2)}\rho_{B}\}$ and so on,

\begin{figure}
\begin{center}
 \includegraphics[width=20pc,height=15pc]{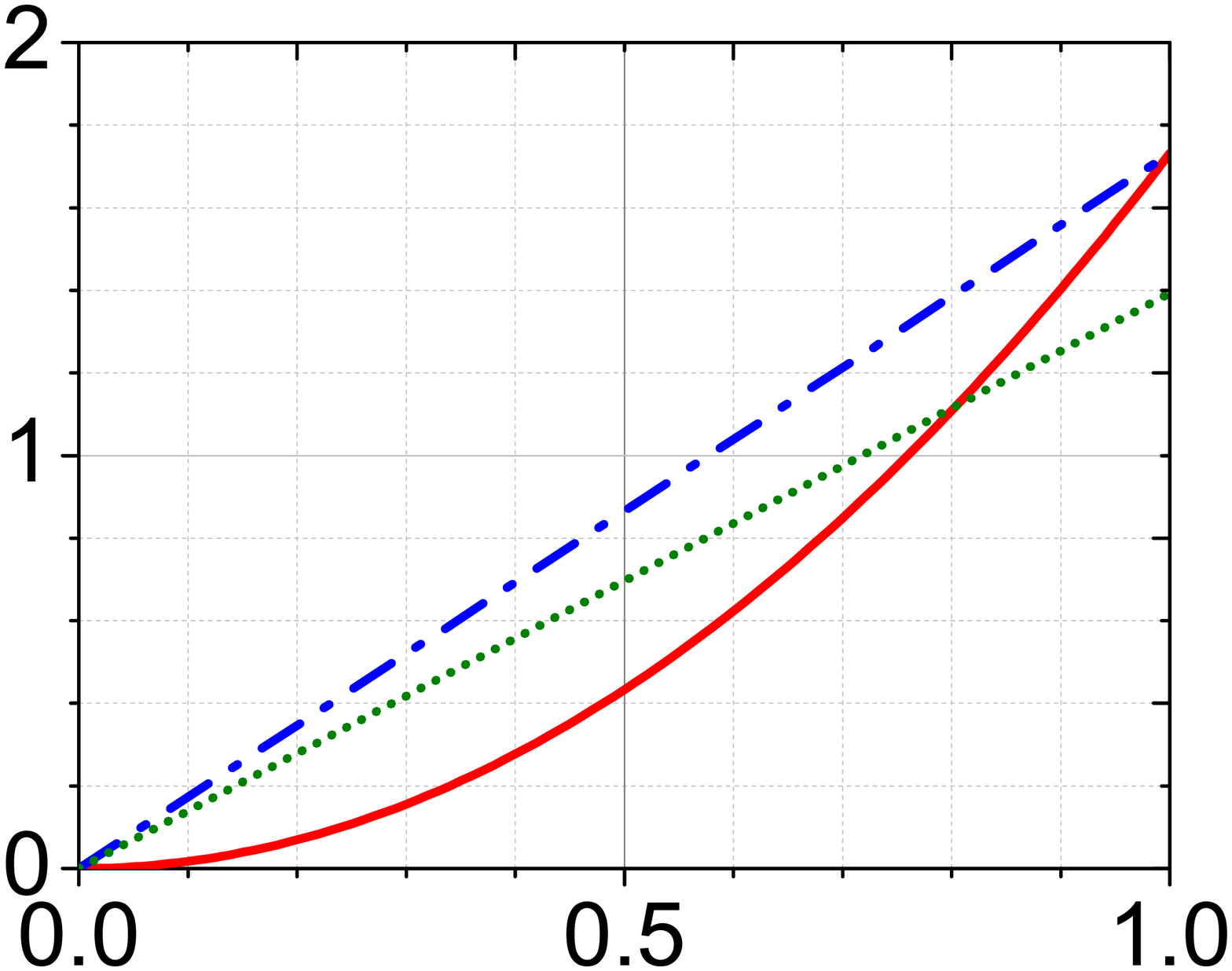}
 \put(-120,0){\Large $x$}
 \put(-245,90){Telp}
 \put(-70,150){$(a)$}
 \includegraphics[width=20pc,height=15pc]{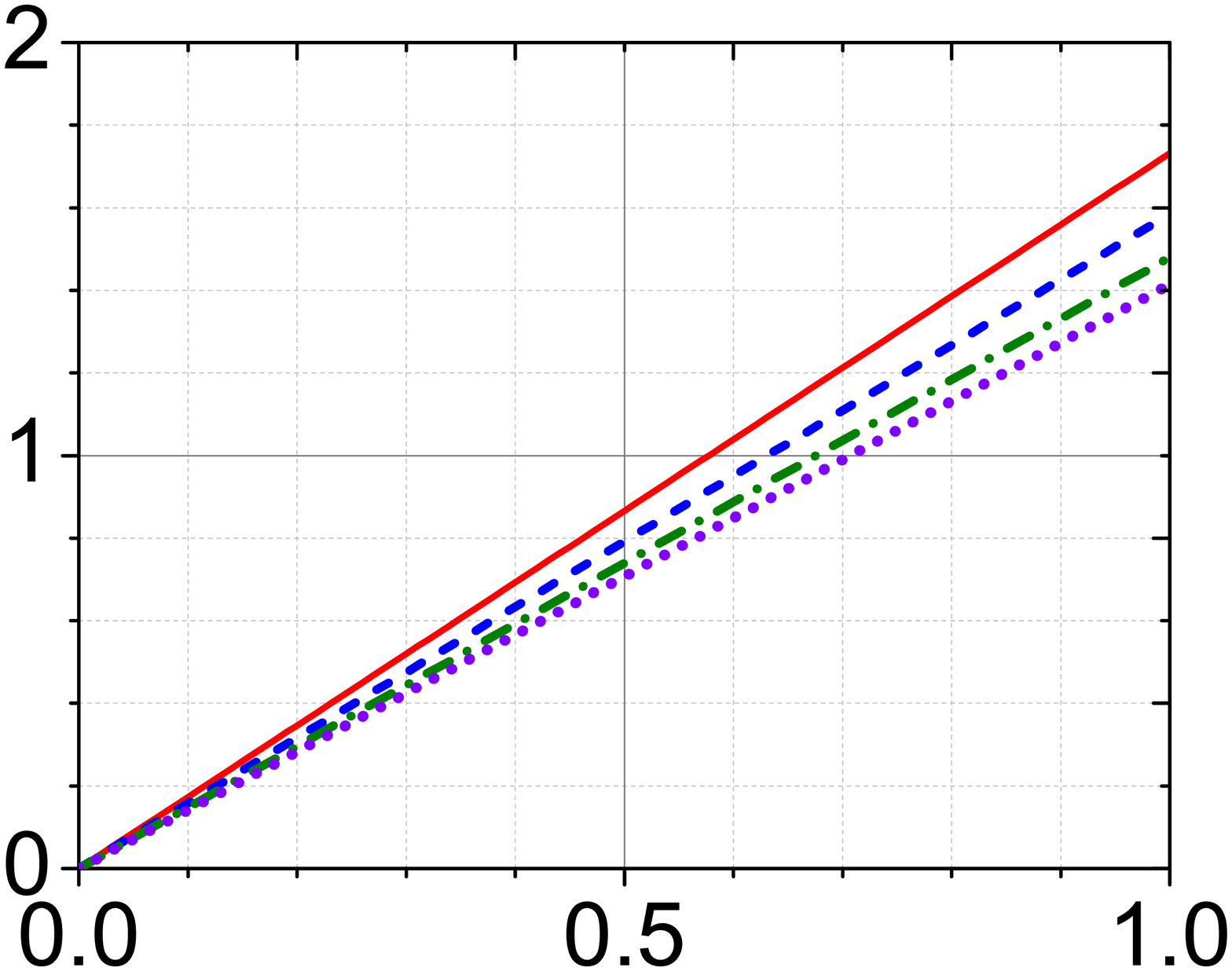}
 \put(-120,0){\Large $x$}
 \put(-245,90){ $Telp$}
   \put(-60,150){$(b)$}
\end{center}
\caption{The efficiency of the  WQBS  to perform quantum
teleportation.(a)The solid, dot and dash-dot for $WW,WX$ and $WB$
bridges, respectively (ba)For $WP$ bridge where, the solid, dash,
dash-dot and dot curves for $p=0,0.7,0.9$,and $1$, respectively.}
\end{figure}

The behavior of   teleportation inequality  $(8)$, is described in
Fig.(4) for different wireless quantum bridges. In Fig.(4a), the
behavior of the teleportation inequality is displayed  for $WW,WX$
and $WB$ bridges. It is clear that, the possibility of using the
generated wireless bridges  as quantum channels  to perform
quantum teleportation, depending on the initial degree  of
entanglement. However, the  $WW$ bridge is useful for quantum
teleportaion for $x\geq 0.8$, while $WX$ bridge for $x\geq 0.6$
and $BW$ bridge for $x\geq 0.5$. Fig.(4b) describes the behavior
of the teleportation inequality (8) for $WP$ bridge.  This figure
shows that, the possibility of using the  $WP$ bridge as quantum
channel increases as $p$ decreases.

 From this figure, we can find the lower values of Werner
 parameter $(x)$, where the generated wireless quantum bridges are
 useful for quantum teleportation. This means that if one can
 improve the efficiency of the source that sends Werner states,
 one can increases the efficiency of the generated wireless
 quantum bridges.

\subsection{Teleportation} In this section, we investigate the
possibility of using the more powerful  wireless quantum bridges
(WQBS) to teleport an unknown quantum signal given by,
\begin{equation}
\rho_u=\left(\begin{array}{cc}
 |\alpha|^2&\alpha\beta^*\\
\alpha^*\beta&|\beta|^2
\end{array}\right), \quad  \mbox{with}\quad
|\alpha|^2+|\beta|^2=1,
\end{equation}
 from one  hop's partners  to another. Let us first consider that, the
 nodes use $XX$ wireless bridge. In this case, the  nodes use the state
(6) as a quantum channel to perform the original protocol
\cite{benn} which is based on local operations followed by Bell
measurements at the sender  hops' node. These measurements are
send via classical channel to the receiver's  hop, who performs
some local operations depending  on the received  classical
information. However, if the sender measures Bell state
$\rho_{\phi^+}$, then the sent quantum signal is retrieved at
receiver's hand  with a fidelity given by,
\begin{eqnarray}
\mathcal{F}_{\mu\nu}&=&|\alpha|^2\Bigl\{(\varrho_{11}+\varrho_{22})+(\alpha\beta^*+\alpha^*\beta)(\varrho_{11}-\varrho_{22})\Bigr\}
\nonumber\\
&&+\alpha^*\beta\bigl\{(|\alpha|^2-|\beta|^2)(\varrho_{14}+\varrho_{23})+
(\alpha\beta^*-\alpha^*\beta)*(\varrho_{14}-\varrho_{23})\Bigr\}
\nonumber\\
&&+\alpha\beta^*\Bigl\{(|\alpha|^2-|\beta|^2)(\varrho_{14}+\varrho_{23})-
(\alpha\beta^*-\alpha^*\beta)*(\varrho_{14}-\varrho_{23})\Bigr\}
\nonumber\\
&&+|\beta|^2\Bigl\{(\varrho_{11}+\varrho_{22})-(\alpha\beta^*+\alpha^*\beta)(\varrho_{11}-\varrho_{22})\Bigr\},
\end{eqnarray}
where $\mu\nu=WW,WB$ or $WX$, if the users use Werener-Werner,
Werner-Bell and X-Wernner bridges, respectively.
\begin{figure}
\begin{center}

 \includegraphics[width=25pc,height=15pc]{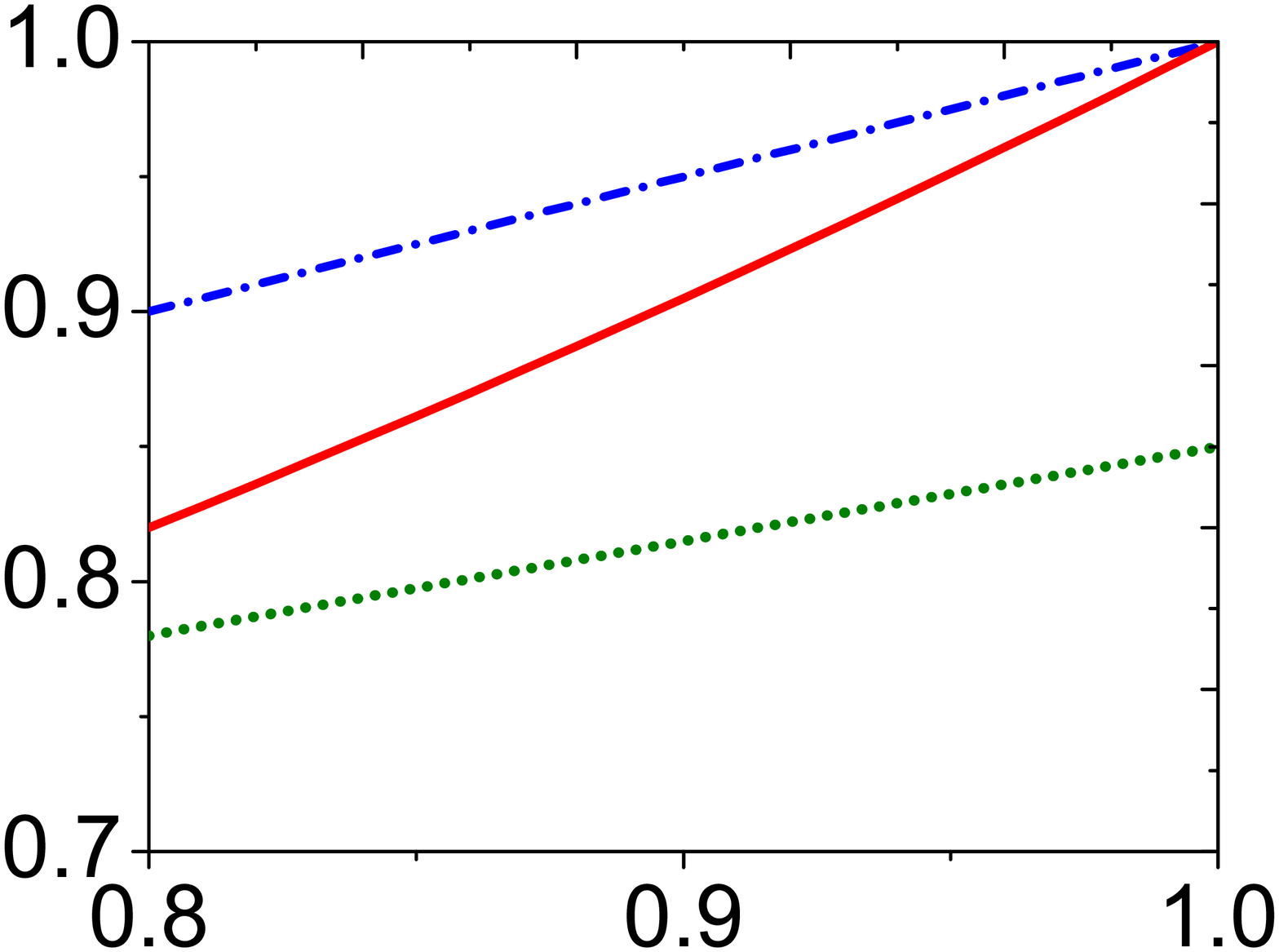}
 \put(-140,0){\Large $x$}
 \put(-300,90){\Large $\mathcal{F}_{\mu\nu}$}
 \end{center}
\caption{ The fidelity of the teleported state (9)  with
$\alpha=\beta=\frac{1}{\sqrt{2}}$. The solid, dash and dot curves
represent the fidelity, $\mathcal{F}_{\mu\nu}$ of the teleported
state by using the wireless bridges $WW$, $WB$ and $WX$(
$c_{xx}=-0.9, c_{yy}=-0.8, c_{zz}=-0.7$), respectively.}
\end{figure}
\\
Fig.(5), describes the behavior of the fidelity of the teleported
quantum signal via the  wireless quantum bridge $WW$
(solid-curve), $WB$ bridge(dash-dot curve) and $WX$  bridge (dot
curve), where we consider  only the bridges which are generated at
$x\geq 0.7$ (efficient bridges for teleportation). It is clear
that, the initial fidelity of the teleported state depends on the
initial entanglement. As an example, if  the partners use the
 $WB$ bridge with($x\geq 0.7)$, the initial fidelity of
the teleported  quantum signal  is large. However, as $x$
increases, the fidelity $\mathcal{F}_{\mu\nu}$ increases to reach
its maximum value ($\mathcal{F}_{\mu\nu}=1$ at $x=1$). On the
other hand, if the partners use the wireless $WW$ bridge, then the
initial fidelity is smaller than the  previous case. As the
Werener's parameter $x$ increases, the fidelity increases to
become maximum at $x=1$, namely, the initially states of the two
hops' members turn into Bell states.
 Finally, the users use the generated $WX$  bridge
then the initially fidelity depends on the degree of entanglement
of $X-$ state. However, the fidelity increases as $x$ increases to
reach its maximum bounds.

\begin{figure}
\begin{center}
 \includegraphics[width=25pc,height=15pc]{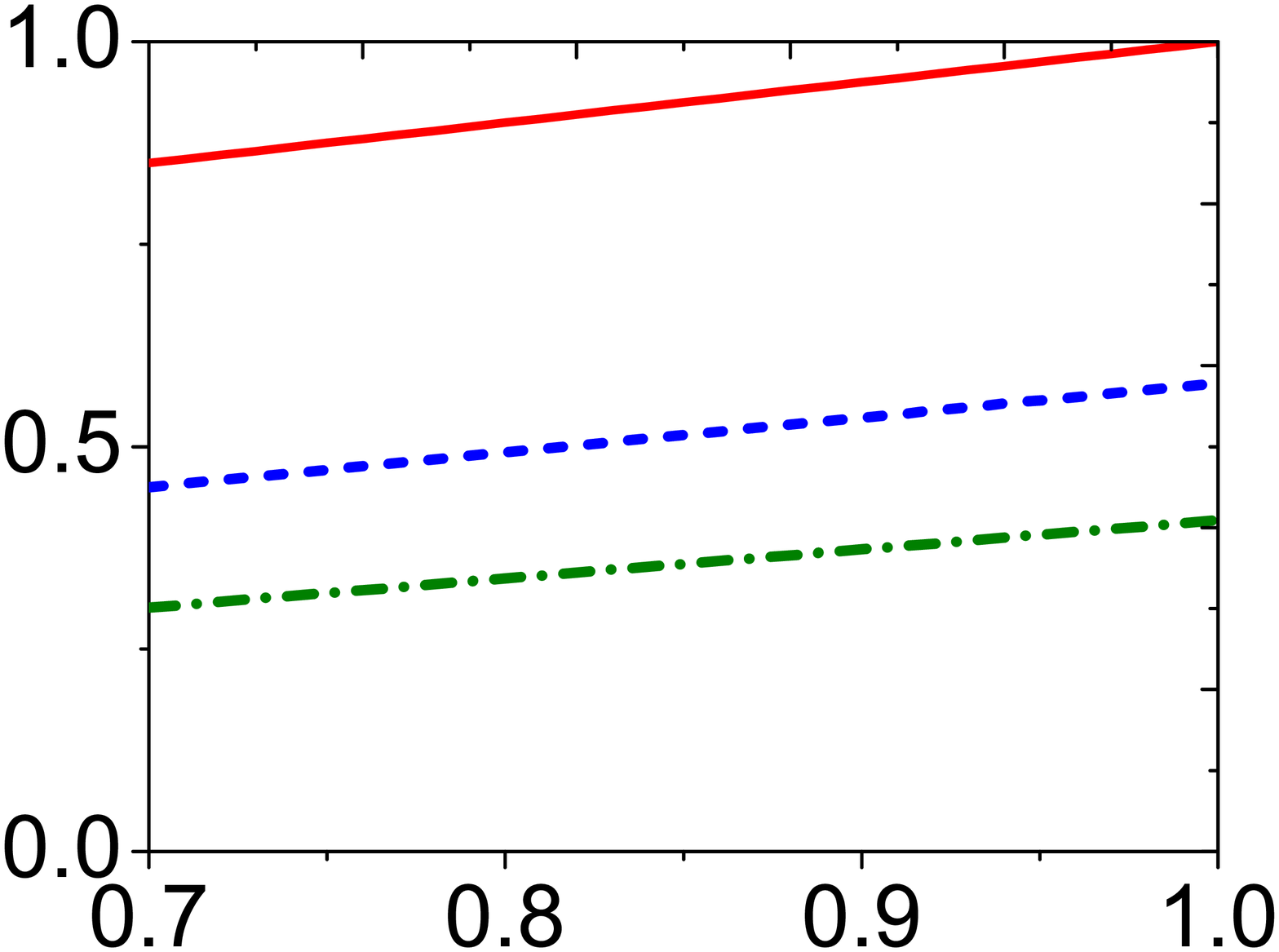}
\put(-140,0){\Large $x$}
 \put(-310,90){\Large $\mathcal{F}_{WP}$}
 \end{center}
\caption{ The  same as Fig.(5), but the users use the wireless
$WP$ bridge. The solid, dash and dash-dot curves for $p=0,0.7$ and
$0.9$, respectively.}
\end{figure}

Finally, if the users decide to use the wireless quantum bridge
$(WP)$ to teleport the unknown quantum signal (9) with a fidelity
given by,
\begin{eqnarray}
\mathcal{F}_{WP}&=&\alpha^2\Bigl\{\tilde\varrho_{11}+\tilde\varrho_{22}+2(\alpha^2-\beta^2)\tilde\varrho_{24}+
(\alpha^*\beta+\beta^*\alpha)(\tilde\varrho_{11}-\tilde\varrho_{22})\Bigr\}
\nonumber\\
&&+(\alpha^*\beta+\beta^*\alpha)\Bigl\{(\alpha^2-\beta^2)(\tilde\varrho_{23}-\tilde\varrho_{14})
-(\alpha^*\beta+\beta^*\alpha)\tilde\varrho_{23}-2\tilde\varrho_{12}\Bigr\}
\nonumber\\
&&+\beta^2\Bigl\{\tilde\varrho_{11}+\tilde\varrho_{22}+2(\alpha^2-\beta^2)\tilde\varrho_{24}-
(\alpha^*\beta+\beta^*\alpha)(\tilde\varrho_{11}-\tilde\varrho_{22})\Bigr\}
\end{eqnarray}

  Fig.(6) shows the behavior of the fidelity $\mathcal{F}_{WP}$.
  The behavior shows that, the initial
fidelity depends on the parameter $p$, where for $p=0$ the pure
state turns into a Bell state. Therefore, the initial fidelity of
the teleported state $\mathcal{F}_{WP}$ is larger. This fidelity
reaches its maximum  value ($\mathcal{F}_{WP}=1$ at $x=1$), which
means that the two  hops share a Bell state. However as $p$
increases, the initial fidelity of the teleported state decreases
and the maximum bounds are reached at $x=1$. The maximum bounds
decreases as $p$ increases.

\section{Purification}
Quantum purification has been used to distill small number of
strongly  entangled qubits  from a large number of weakly
entangled qubits, via local operations, classical communication
and measurements. The first purification protocol (IBM) has
proposed by Bennett et al.\cite{Bennt96}, where they obtain the
singled states from Werner classes. Deutsch et al. \cite{Deutch}
have suggested the  Oxford protocol which is more efficient than
IBM protocol. Since then there are several protocols have been
suggested. For example, a more efficient entanglement purification
protocol is suggested by Metwally\cite{Metwally02}, which is more
efficient than the IBM and Oxford protocols. Another improvement
has been done on the IBM protocol by Feng et al.\cite{Feng}. All
the previous protocol have been improved  by several versions.
Among of these improvements the protocol which is introduced by
Metwally and Obada \cite{Metwally06}, where this improved version
based on using  the controlled-controlled NOT gate (CCNOT) instead
of CNOT.

In this context, we can use one of the previous protocols to
distill a wireless quantum bridges with high degree of
entanglement from weakly entangled bridges. In this wireless
quantum network, we suggested two strategies: The first  is the
initial  partial entangled state can be purified before sending
them to the
 hops'nodes. In this case, all the users will be supplied by
MES, and the protocol turns into Wang et al. protocol \cite{Wang}.
The second possibility is  performing a quantum purification
protocol on the less entangled  bridges (useless bridges for
teleportation) to increase their efficiency. However this will be
our next contribution to find which strategy is better.

\section{Conclusion}
The possibility of  generating wireless quantum networks (WQNS)
between different  hops'nodes, where it is assumed that these
nodes share together arbitrary two qubit systems randomly, is
discussed. To achieve quantum communication between the
non-connected  hops'nodes, the users have to generated  wireless
quantum bridges. The type of  these Wireless bridges  depends on
the states which are shared between the terminals of each  hop,
where we have generated Werner-Werner, Werner-Bell,Werner- $X$ and
Werner-Pure bridges. The entanglement of each WQB is quantified by
the means of concurrence. It is shown that, for less entangled
state the non-connected  hops' nodes  turn into  wireless  quantum
bridges for larger values of Werener's parameter, $x$. However,
the partial entangled wireless quantum  bridges turn into a
maximum entangled  wireless bridges  when the Werner's parameter
namely ($x=1$). The entanglement of the Werner-Pure (WP) bridges,
depends on pure  and Werner states's parameters, where it
increases for larger  values of Werner parameter and smaller
values of the pure state  parameter.

The efficiency of the generated wireless  quantum bridges to
perform quantum teleportation is discussed for different types of
bridges. It is shown that, the teleportation inequality is
violated for small values of  the Werner's parameter and
consequently the efficiency of the WQBS to perform quantum
teleportation  decreases. However, this efficiency of the
generated wireless quantum bridges increase for larger values of
Werner's parameter and smaller values of the pure state's
parameter.

The more powerful wireless quantum bridges are used to teleport
unknown  quantum signals from  one node to another, where we
consider only the bridges which  obey the teleportation
inequality. The fidelity of the teleported  quantum signal
increases by increasing Werner's parameter or decreasing the pure
state parameter for $WP$ bridge. The maximum value of the fidelity
depends on the entanglement of the used wireless quantum bridge.

{\it In conclusion:} a wireless quantum networks (WQNS) can be
generated between different  hops' nodes  sharing arbitrary
different two qubit states. The efficiency of the WQN and hence
its ability to perform wireless quantum communication can be
enhanced by controlling  the devices which generate these signals.

\end{document}